\documentclass[11pt]{amsart}
\usepackage{hyperref}
\theoremstyle{definition}

\theoremstyle{remark}

\numberwithin{equation}{section}

\def \dd{{\rm d}}
\begin{document}
\title[Wave propagation in Einstein's unified field theory]
{The propagation of waves in Einstein's unified field theory
as shown by two exact solutions}%
\author{Salvatore Antoci}%
\address{Dipartimento di Fisica ``A. Volta'' and IPCF of CNR, Pavia, Italy}%
\email{Antoci@fisicavolta.unipv.it}%

\begin{abstract}
The propagation of waves in two space dimensions exhibited by two
exact solutions to the field equations of Einstein's unified field
theory is investigated under the assumption that the metric
$s_{ik}$ is the one already chosen by Kur\c{s}uno\u{g}lu and by
H\'ely in the years 1952-1954. It is shown that, for both exact
solutions, with this choice of the metric the propagation of the
waves occurs in the wave zone with the fundamental velocity ($\dd
s^2=0$).
\end{abstract}
\maketitle
\section{Introduction}\label{S1}
As soon as the independent, but concurrent efforts by Einstein
 and by Schr\"odinger eventually led to the final mathematical formulation
of what may be respectively called the metric-affine
\cite{Einstein1925, ES1946, Einstein1948, EK1955} and the purely
affine \cite{Schroedinger1947a, Schroedinger1947b,
Schroedinger1948, Schroedinger1951} versions for the nonsymmetric
generalization of Einstein's theory of 1915, skilled theoreticians
and geometers \footnote[1]{Without pretense of completeness, let
us here recall the remarkable achievements by Papapetrou
\cite{Papapetrou1948}, Wyman \cite{Wyman1950}, Kur\c{s}uno\u{g}lu
\cite{Kursunoglu1952a, Kursunoglu1952b}, Lichnerowicz
\cite{Lichnerowicz1954, Lichnerowicz1955},  H\'ely
\cite{Hely1954a, Hely1954b}, Tonnelat \cite{Tonnelat1955}, V.V.
Narlikar and B.R. Rao \cite{Narlikar Rao1956}, Treder
\cite{Treder1957}, Hlavat\'y \cite{Hlavaty1957}.} undertook the
difficult task of understanding the physical meaning of the theory
through the investigation of its mathematical structure and the
search for the solutions, both exact and approximate, to its field
equations.\par However, progress towards the accomplishment of
this task was very slow, if in 1954 Schr\"odinger still wrote
\cite{Schroedinger1954} about the very identification of the
metric tensor of the theory as an open question \footnote[2]{In
the last pages of the cited book, he wrote in fact: ``We cannot
even feel sure whether in the nonsymmetric case the $g_{(ik)}$ or
the ${\mathbf g}^{(ik)}$ (or, less likely, the ${\mathbf
g}_{(ik)}$ or the $g^{(ik)}$) play the part of the corresponding
tensorial entities describing the gravitational field in
Einstein's theory''. For the definitions of the quoted quantities,
see appendix (\ref{A}) of the present paper.}. A quite relevant
contribution to the identification of the metric tensor in
Einstein-Schr\"odinger unified field theory came from the study of
the Cauchy problem done by Lichnerowicz \cite{Lichnerowicz1954,
Lichnerowicz1955}. He succeeded in thorougly analysing the Cauchy
problem without solving the unwieldy equation (\ref{A.2})
explicitly,  and proved that the answer to the Cauchy problem is
in general unique, unless the surface $S$, on which the Cauchy
data are given, is a characteristic surface, i.e. unless locally
$S=f(x^i)$, where $f$ fullfils the so-called eikonal equation
\begin{equation}\label{1.1}
g^{(ik)}{f_{,i}}{f_{,k}}=0.
\end{equation}
Although, from a purely mathematical perspective, the function $f$
satisfying (\ref{1.1}) only defines a surface that is unsuitable
as a startpoint for the solution of the Cauchy problem, the very
fact that (\ref{1.1}) has just the form of the eikonal equation,
i.e. of the equation that stems from d'Alembert equation in the
high frequency limit, naturally led to read in it a law of wave
propagation, sometimes of a shock wave propagation, ruled by a
metric, in the present case by $g^{(ik)}$, or, more precisely, by
any tensor conformally related to it. Indeed ${\mathbf g}^{(ik)}$,
one of the four candidates considered by Schr\"odinger
\cite{Schroedinger1954} for producing a metric, through the
stipulation
\begin{equation}\label{1.2}
\sqrt{-s}s^{ik}=\mathbf{g}^{(ik)},
\end{equation}
where $s=\det{(s_{ik})}$, allows defining a metric tensor $s^{ik}$
that is conformally related to $g^{(ik)}$. Why, among all the
tensors that are conformally related to $g^{(ik)}$, just $s^{ik}$
should be chosen as metric, turns out from the quoted results
found by Kur\c{s}uno\u{g}lu \cite{Kursunoglu1952a,
Kursunoglu1952b} and by H\'ely \cite{Hely1954a}: with that choice
the four identities, that render the field equations (\ref{A.2}) -
(\ref{A.5}) compatible, assume a very simple and allusive writing.
So allusive that, on this basis, H\'ely \cite{Hely1954b} decided
to disobey the injunction both by Einstein and by Schr\"odinger,
according to which no phenomenological source terms should be
appended at the right-hand sides of the field equations
(\ref{A.2}) - (\ref{A.5}). In that way the conservation
identities, that are otherwise empty, appear to assume physical
meaning. More recently, while retaining H\'ely's choice of the
metric tensor $s_{ik}$, and by availing of a crucial finding by
Borchsenius \cite{Borchsenius1978}, H\'ely's approach was extended
\cite{Antoci1991}, by adding phenomenological sources at the
right-hand sides of all the field equations, in the form of a
symmetric energy-momentum tensor, and of two conserved
four-current densities. The way for achieving this result, for the
reader's convenience, is recalled in appendix (\ref{B}).\par It is
clear, however, that the assumption that $s_{ik}$, as defined by
(\ref{1.2}), can be the metric of Einstein-Schr\"odinger theory,
is an hypothesis that needs further confirmation. Just the
retrieval and the study of exact solutions of (\ref{A.2}) -
(\ref{A.5}) displaying a wavy behaviour can either confirm or
disprove the hypothesis \footnote[3]{Needless to say, while a
confirmation would be always provisional, a disproval would be a
definitive one.}. Happily enough, two such solutions do exist.
They belong to the class of exact solutions intrinsically
depending on three coordinates \cite{Antoci1987}, whose structure
is recalled in appendix (\ref{C}).
\section{Wave propagation in two exact solutions}\label{S2}
An exact solution allowing for wave propagation in two space and
one time dimensions is easily built by the method of appendix
(\ref{C}), provided that one chooses
\begin{equation}\label{2.1}
h_{ik}=\eta_{ik}\equiv\text{diag}(-1,-1,-1,1)
\end{equation}
as seed metric for the Hermitian solution, defined with respect to
the coordinates $x^1=x$, $x^2=y$, $x^3=z$, $x^4=t$. In these
coordinates the fundamental form $g_{ik}$ of the mentioned
solution reads:
\begin{equation}\label{2.2}
g_{ik}=\left(\begin{array}{rrrr}
 -1 &  0 &  e & 0 \\
  0 & -1 &  f & 0 \\
 -e & -f &  v & c \\
  0 &  0 & -c & 1
\end{array}\right),
\end{equation}
with
\begin{equation}\label{2.3}
v=-1-c^2+e^2+f^2
\end{equation}
and
\begin{equation}\label{2.4}
e=i\xi_{,x}, \ f=i\xi_{,y}, \ c=-i\xi_{,t}, \ \ i=\sqrt{-1},
\end{equation}
where the function $\xi=\xi(x,y,t)$ fullfils, in the chosen
representative space, the d'Alembert equation
\begin{equation}\label{2.5}
\xi_{,xx}+\xi_{,yy}-\xi_{,tt}=0
\end{equation}
with respect to the three coordinates $x$, $y$, $t$. When $\xi$ is
defined by (\ref{2.4}), besides the field equation (\ref{A.3}),
also the unsolicited, invariant equation
\begin{equation}\label{2.6}
g_{[ik],l}+g_{[kl],i}+g_{[li],k}=0
\end{equation}
is satisfied, i.e. the antisymmetric field $g_{[ik]}$ appears to
be endowed with electromagnetic meaning \footnote[4]{Other
solutions fulfilling (\ref{2.6}), and representing the general
electrostatic solution \cite{ALM2005} and the magnetic field
generated by constant electric currents running on $n$ parallel
wires \cite{Antoci2009} have been previously investigated by using
$s_{ik}$ as metric.}.\par If the metric $s_{ik}$, defined by eq.
(\ref{1.1}), were equal to the seed metric $h_{ik}$ defined by
(\ref{2.1}), the solution would represent electromagnetic waves
that propagate with the fundamental velocity ($\dd s^2=0$). The
interval of the chosen metric $s_{ik}$, however, differs from the
Minkowski interval. It is defined by equation (\ref{C.13}) that,
in the case of the particular solution defined by (\ref{2.2}) -
(\ref{2.4}), reads
\begin{equation}\label{2.7}
\dd s^2=s_{ik}\dd x^{i}\dd x^{k}=\sqrt{-v}(\dd t^2-\dd x^2-\dd
y^2- \dd z^2)+\frac{(\dd \xi)^2}{\sqrt{-v}}.
\end{equation}
If the second term at the right hand side of (\ref{2.7}) were
absent, the propagation of the electromagnetic waves would occur
with the fundamental velocity also with respect to the chosen
metric $s_{ik}$, because the first term at the right hand side is
just conformally related to the square of the Minkowski interval.
But a moment's reflection shows that, when $\xi$ has, in the
``Bildraum'', a truly wave zone behaviour, hence a ``Bildraum''
wavevector can be locally defined, $\dd\xi$, when taken along the
direction of that wavevector, necessarily vanishes. As a
consequence $\dd s^2$, as defined by (\ref{2.7}), shall vanish in
the direction of the ``Bildraum'' wavevector. One concludes that,
with respect to the chosen metric $s_{ik}$, in the considered
electromagnetic solution the electromagnetic waves in the wave
zone do propagate with the fundamental velocity\footnote[5]{The
proof, that  in the wave zone of the considered electromagnetic
solution the electromagnetic waves do propagate with the
fundamental velocity in the metric sense ($\dd s^2=0$), is briefly
outlined here, by availing of the very well known properties of
D'Alembert's equation in the chosen ``Bildraum''. In a small
neighbourhood of the wave zone, by suitable choice of the
coordinates (otherwise, by suitable choice of the particular
solution) equation (\ref{2.5})
 can be reduced to
\begin{equation}\nonumber
\xi_{,xx}-\xi_{,tt}=0.
\end{equation}
In that small neighbourhood of the wave zone, a particular
solution reads, say
\begin{equation}\nonumber
\xi=\xi(x-t),
\end{equation}
for which
\begin{equation}\nonumber
\dd\xi=0
\end{equation}
when
\begin{equation}\nonumber
\dd x=\dd t,
\end{equation}
as needed. Therefore, since $\dd y=\dd z=0$, the interval $\dd
s^2$ defined by (\ref{2.7}) vanishes locally in the wave zone.} in
the metric sense ($\dd s^2=0$).\par

Another exact solution, endowed with axial symmetry, and allowing
too for wave propagation in two space and one time dimensions, is
built by the same method of appendix (\ref{C}), provided that one
now chooses
\begin{equation}\label{2.8}
h_{ik}=\text{diag}(-1,-1,-r^2,1),
\end{equation}
defined with respect to polar cylindrical coordinates $x^1=r$,
$x^2=z$, $x^3=\varphi$, $x^4=t$, as seed metric for the Hermitian
solution. Its fundamental tensor $g_{ik}$ reads:
\begin{equation}\label{2.9}
g_{ik}=\left(\begin{array}{rrrr}
  -1 & 0 & \delta & 0 \\
  0 & -1 & \varepsilon & 0 \\
  -\delta & -\varepsilon & \zeta & \tau \\
  0 & 0 & -\tau & ~1
\end{array}\right),
\end{equation}
with
\begin{equation}\label{2.10}
\zeta=-r^2+\delta^2+\varepsilon^2-\tau^2,
\end{equation}
and
\begin{equation}\label{2.11}
\delta=ir^2\psi_{,r}, \ \varepsilon=ir^2\psi_{,z}, \
\tau=-ir^2\psi_{,t},
\end{equation}
where $\psi(r,z,t)$ now fullfils d'Alembert equation in
cylindrical coordinates, namely:
\begin{equation}\label{2.12}
\psi_{,rr}+\frac{\psi_{,r}}{r}+\psi_{,zz}-\psi_{,tt}=0.
\end{equation}
The metric $s_{ik}$ of this solution can be written as
\begin{eqnarray}\label{2.13}
s_{ik}=\frac{\sqrt{-\zeta}}{r} \left(\begin{array}{rrcr}
 -1 &  0 &   0  & 0 \\
  0 & -1 &   0  & 0 \\
  0 &  0 & -r^2 & 0 \\
  0 &  0 &   0  & 1
\end{array}\right)\\\nonumber
+\frac{r^3}{\sqrt{-\zeta}} \left(\begin{array}{cccc}
 \psi_{,r}\psi_{,r} & \psi_{,r}\psi_{,z} & 0 &  \psi_{,r}\psi_{,t} \\
 \psi_{,r}\psi_{,z} & \psi_{,z}\psi_{,z} & 0 &  \psi_{,z}\psi_{,t} \\
          0         &          0         & 0 &           0         \\
 \psi_{,r}\psi_{,t} & \psi_{,z}\psi_{,t} & 0 &  \psi_{,t}\psi_{,t} \\
 \end{array}\right),
\end{eqnarray}
hence the square of the line element, in the adopted coordinates,
comes to read
\begin{equation}\label{2.14}
\dd s^2=s_{ik}\dd x^{i}\dd x^{k}
=\frac{\sqrt{-\zeta}}{r}\left(-\dd r^2-\dd z^2-r^2\dd
{\varphi}^2+\dd t^2\right) +\frac{r^3}{\sqrt{-\zeta}}(\dd\psi)^2.
\end{equation}
This solution has nothing to do with Maxwell's equations, because
with the seed metric (\ref{2.8}) the additional conditions
(\ref{C.3}) no longer have any relation to the electromagnetic
looking equation (\ref{2.6}).\par A particular, time independent
solution, obtained too from the same seed (\ref{2.8}), for which
\begin{equation}\label{2.15}
\psi=-\sum_{q=1}^n K_q\ln\frac{p_q+z-z_q}{r}, \ \
\end{equation}
where
\begin{equation}\label{2.16}
p_q=[r^2+(z-z_q)^2]^{1/2},
\end{equation}
while $K_q$ and $z_q$ are constants, has been investigated
\cite{ALM2006} some time ago. In keeping with an earlier
approximate calculation done by Treder \cite{Treder1957}, it
proves that pole sources at rest, defined by eq. (\ref{B.15}),
interact with forces not depending on their mutual distance, like
the quarks of chromodynamics are supposed to do. The axially
symmetric waves that we are presently considering should be
therefore emitted and absorbed by such pole sources.\par Whatever
their physical meaning, the velocity with which these waves
propagate is easily ascertained, like it occurred with the
electromagnetic example considered previously. In fact, if the
second term at the right hand side of (\ref{2.14}) were absent,
i.e. when $\dd \psi$ is vanishing, the squared interval $\dd s^2$,
referred to cylindrical coordinates, would be conformally
Minkowskian, and the speed of propagation of a wave with respect
to the metric $s_{ik}$ should be equal to the fundamental velocity
that prevails with respect to the seed metric (\ref{2.8}). But,
again, a moment's reflection shows that, when $\psi$ has, in the
``Bildraum'', a truly wave zone behaviour, hence when a
``Bildraum'' wavevector can be locally defined, $\dd\psi$
necessarily vanishes, when taken along the direction of that
wavevector. As a consequence $\dd s^2$, as defined by
(\ref{2.14}), shall vanish in the direction of the ``Bildraum''
wavevector. Therefore, with the chosen metric $s_{ik}$, in the
considered solution the waves of $g_{[ik]}$, whatever their
physical meaning, do propagate in the wave zone with the
fundamental velocity\footnote[6]{A proof closely similar to the
one given in the previous footnote applies here, and is left to
the ingenuity of the reader.} in the metric sense ($\dd
s^2=0$).\par In both the considered examples, the choice of the
metric $s_{ik}$ done by Kur\c{s}uno\u{g}lu \cite{Kursunoglu1952a,
Kursunoglu1952b} and by H\'ely \cite{Hely1954a, Hely1954b} happens
therefore to be compatible with the wavy behaviour exhibited by
the exact solutions.

\appendix
\section{Einstein's unified field theory; Hermitian version}\label{A}
We consider here the Hermitian version \cite{Einstein1948} for
Einstein's nonsymmetric generalization of the theory of 1915. A
given geometric quantity \cite{Schouten1954} is called Hermitian
with respect to the indices $i$ and $k$, both either covariant or
contravariant, if the part of the quantity that is symmetric with
respect to $i$ and $k$ is real, while the part that is
antisymmetric is purely imaginary. We contemplate the Hermitian
fundamental form $g_{ik}=g_{(ik)}+g_{[ik]}$, and the affine
connection $\Gamma^i_{kl}=\Gamma^i_{(kl)}+\Gamma^i_{[kl]}$,
Hermitian with respect to the lower indices; both entities depend
on the real coordinates $x^i$, while $i$ runs from 1 to 4. We
define also the Hermitian contravariant tensor $g^{ik}$ through
the relation
\begin{equation}\label{A.1}
g^{il}g_{kl}\equiv g^{li}g_{lk}=\delta^i_k,
\end{equation}
and the contravariant tensor density $\mathbf{
g}^{ik}=(-g)^{1/2}g^{ik}$; $g\equiv\det{(g_{ik})}$ is a real
quantity. Then the field equations of Einstein's unified field
theory in the complex Hermitian form \cite{Einstein1948} come to
read:
\begin{eqnarray}\label{A.2}
g_{ik,l}-g_{nk}\Gamma^n_{il}-g_{in}\Gamma^n_{lk}=0,\\\label{A.3}
\mathbf{ g}^{[is]}_{~~,s}=0,\\\label{A.4}
R_{(ik)}(\Gamma)=0,\\\label{A.5}
R_{[ik],l}(\Gamma)+R_{[kl],i}(\Gamma)+R_{[li],k}(\Gamma)=0;
\end{eqnarray}
$R_{ik}(\Gamma)$ is the Hermitian Ricci tensor
\begin{equation}\label{A.6}
R_{ik}(\Gamma)=\Gamma^a_{ik,a}-\Gamma^a_{ia,k}
-\Gamma^a_{ib}\Gamma^b_{ak}+\Gamma^a_{ik}\Gamma^b_{ab}.
\end{equation}

\section{Adding phenomenological sources to all the Hermitian field
equations of Einstein}\label{B}
In a four-dimensional manifold,
let $\mathbf{g}^{ik}$ be a contravariant tensor density with an
even part $\mathbf{g}^{(ik)}$ and an alternating one
$\mathbf{g}^{[ik]}$:
\begin{equation}\label{B.1}
\mathbf{g}^{ik}=\mathbf{g}^{(ik)}+\mathbf{g}^{[ik]},
\end{equation}
and $W^i_{kl}$ be a general affine connection
\begin{equation}\label{B.2}
W^i_{kl}=W^i_{(kl)}+W^i_{[kl]}.
\end{equation}
For the Riemann curvature tensor built from $W^i_{kl}$:
\begin{equation}\label{B.3}
R^i_{~klm}(W)=W^i_{kl,m}-W^i_{km,l}
-W^i_{al}W^a_{km}+W^i_{am}W^a_{kl},
\end{equation}
two distinct contractions exist, $R_{ik}(W)=R^p_{~ikp}(W)$ and
$A_{ik}(W)=R^p_{~pik}(W)$ \cite{Schroedinger1954}. But the
transposed affine connection $\tilde{W}^i_{kl}=W^i_{lk}$ must be
considered too: from it, the Riemann curvature tensor
$R^i_{~klm}(\tilde{W})$ and its two contractions
$R_{ik}(\tilde{W})$ and $A_{ik}(\tilde{W})$ can be formed as well.
We aim at following the pattern of the general relativity of 1915,
which is built by the variational method from the Lagrangian
density $\mathbf{g}^{ik}R_{ik}$, but now any linear combination
$\bar{R}_{ik}$ of the four above-mentioned contractions is
possible. A good choice \cite{Borchsenius1978}, for physical
reasons that will become apparent later, is
\begin{equation}\label{B.4}
\bar{R}_{ik}(W)=R_{ik}(W)+\frac{1}{2}A_{ik}(\tilde{W}).
\end{equation}
Let us provisionally endow the theory with sources in the form of
a nonsymmetric tensor $P_{ik}$ and of a current density
$\mathbf{j}^i$, coupled to $\mathbf{g}^{ik}$ and to the vector
$W_i=W^l_{[il]}$ respectively. The Lagrangian density
\begin{equation}\label{B.5}
\mathbf{L}=\mathbf{g}^{ik}\bar{R}_{ik}(W)
-8\pi\mathbf{g}^{ik}P_{ik} +\frac{8\pi}{3}W_i\mathbf{j}^i
\end{equation}
is thus arrived at.  By performing independent variations of the
action $\int\mathbf{L}d\Omega$ with respect to $W^p_{qr}$ and to
$\mathbf{g}^{ik}$ with suitable boundary conditions we obtain the
field equations
\begin{eqnarray}\label{B.6}
-\mathbf{g}^{qr}_{~,p}+\delta^r_p\mathbf{g}^{(sq)}_{~,s}
-\mathbf{g}^{sr}W^q_{sp}-\mathbf{g}^{qs}W^r_{ps}\\\nonumber
+\delta^r_p\mathbf{g}^{st}W^q_{st} +\mathbf{g}^{qr}W^t_{pt}
=\frac{4\pi}{3}(\mathbf{j}^r\delta^q_p-\mathbf{j}^q\delta^r_p)
\end{eqnarray}
and
\begin{equation}\label{B.7}
\bar{R}_{ik}(W)=8\pi P_{ik}.
\end{equation}
By contracting eq. (\ref{B.6}) with respect to $q$ and $p$ we get
\begin{equation}\label{B.8}
\mathbf{g}^{[is]}_{~,s}={4\pi}\mathbf{j}^i.
\end{equation}
The very finding of this physically welcome equation entails
however that we cannot determine the affine connection $W^i_{kl}$
uniquely in terms of $\mathbf{g}^{ik}$: (\ref{B.6}) is invariant
under the projective transformation
${W'}^i_{kl}=W^i_{kl}+\delta^i_k\lambda_l$, where $\lambda_l$ is
an arbitrary vector field. Moreover eq. (\ref{B.7}) is invariant
under the transformation
\begin{equation}\label{B.9}
{W'}^i_{kl}=W^i_{kl}+\delta^i_k\mu_{,l}
\end{equation}
where $\mu$ is an arbitrary scalar. By following Schr\"odinger
\cite{Schroedinger1948,Schroedinger1954}, we write
\begin{equation}\label{B.10}
W^i_{kl}=\Gamma^i_{kl}-\frac{2}{3}\delta^i_kW_l,
\end{equation}
where $\Gamma^i_{kl}$ is another affine connection, by definition
constrained to yield $\Gamma^l_{[il]=0}$. Then eq. (\ref{B.6})
becomes
\begin{equation}\label{B.11}
\mathbf{g}^{qr}_{~,p}+\mathbf{g}^{sr}\Gamma^q_{sp}+\mathbf{g}^{qs}\Gamma^r_{ps}
-\mathbf{g}^{qr}\Gamma^t_{(pt)}
=\frac{4\pi}{3}(\mathbf{j}^q\delta^r_p-\mathbf{j}^r\delta^q_p)
\end{equation}
that allows one to determine $\Gamma^i_{kl}$ uniquely, under very
general conditions \cite{Tonnelat1955, Hlavaty1957}, in terms of
$\mathbf{g}^{ik}$. When eq. (\ref{B.10}) is substituted in eq.
(\ref{B.7}), the latter comes to read
\begin{eqnarray}\label{B.12}
\bar{R}_{(ik)}(\Gamma)=8\pi P_{(ik)}\\\label{B.13}
\bar{R}_{[ik]}(\Gamma) =8\pi P_{[ik]}-\frac{1}{3}(W_{i,k}-W_{k,i})
\end{eqnarray}
after splitting the even and the alternating parts. Wherever the
source term is nonvanishing, a field equation loses its r\^ole,
and becomes a definition of some property of matter in terms of
geometrical entities; it is quite obvious that such a definition
must be unique. This occurs with eqs. (\ref{B.8}), (\ref{B.11})
and (\ref{B.12}), but it does not happen for eq. (\ref{B.13}).
This equation only prescribes that $\bar{R}_{[ik]}(\Gamma)-8\pi
P_{[ik]}$ is the curl of the arbitrary vector $W_i/3$; it is
therefore equivalent to the four equations
\begin{equation}\label{B.14}
\bar{R}_{[ik],l}(\Gamma)+\bar{R}_{[kl],i}(\Gamma)+\bar{R}_{[li],k}(\Gamma)
=8\pi\{P_{[ik],l}+P_{[kl],i}+P_{[li],k}\},
\end{equation}
that cannot specify $P_{[ik]}$ uniquely. We therefore scrap the
redundant tensor $P_{[ik]}$, like we scrapped the redundant affine
connection $W^i_{kl}$ of eq. (\ref{B.6}), and assume that matter
is described by the symmetric tensor $P_{(ik)}$, by the conserved
current density $\mathbf{j}^i$ and by the conserved current
\begin{equation}\label{B.15}
K_{ikl}=\frac{1}{8\pi}\{\bar{R}_{[ik],l}+\bar{R}_{[kl],i}+\bar{R}_{[li],k}\}.
\end{equation}
The general relativity of 1915, to which the present theory
reduces when $\mathbf{g}^{[ik]}=0$, suggests rewriting eq.
(\ref{B.12}) as
\begin{equation}\label{B.16}
\bar{R}_{(ik)}(\Gamma)=8\pi(T_{ik} -\frac{1}{2}s_{ik}s^{pq}T_{pq})
\end{equation}
where $s_{ik}=s_{ki}$ is the still unchosen metric tensor of the
theory, $s^{il}s_{kl}=\delta^i_k$, and the symmetric tensor
$T_{ik}$ will act as energy tensor. If, in keeping with the choice
done by Kur\c{s}uno\u{g}lu and by H\'ely, $s_{ik}$ is defined like
in equation (\ref{1.2}), equation (\ref{B.16}) is readily seen to
stem directly from the variation of the Lagrangian (\ref{B.5}),
with a slightly reworked matter term, with respect to the chosen
metric $s_{ik}$.\par
     When sources are vanishing, equations (\ref{B.11}), (\ref{B.16}), (\ref{B.8}) and
(\ref{B.15}) reduce to the original equations of Einstein's
unified field theory, reported in appendix (\ref{A}), because then
$\bar{R}_{ik}(\Gamma)$=${R}_{ik}(\Gamma)$; moreover they enjoy the
property of transposition invariance also when the sources are
nonvanishing. If $\mathbf{g}^{ik}$, $\Gamma^i_{kl}$,
$\bar{R}_{ik}(\Gamma)$ represent a solution with the sources
$T_{ik}$, $\mathbf{j}^i$ and $K_{ikl}$, the transposed quantities
$\tilde{\mathbf{g}}^{ik}=\mathbf{g}^{ki}$,
$\tilde{\Gamma}^i_{kl}=\Gamma^i_{lk}$ and
$\bar{R}_{ik}(\tilde{\Gamma})$= $\bar{R}_{ki}(\Gamma)$ represent
another solution, endowed with the sources $\tilde{T}_{ik}=T_{ik},
\tilde{\mathbf{j}}^i=-\mathbf{j}^i$ and
$\tilde{K}_{ikl}=-K_{ikl}$. Such a physically desirable outcome is
a consequence of the choice made \cite{Borchsenius1978} for
$\bar{R}_{ik}$. These equations intimate that Einstein's unified
field theory with sources should be interpreted like a
gravoelectrodynamics in a polarizable continuum, allowing for both
electric and magnetic currents. The study of the conservation
identities confirms this idea \cite{Antoci1991} and strengthens at
the same time the identification of the metric tensor $s_{ik}$
done by Kur\c{s}uno\u{g}lu and by H\'ely. Let us consider the
invariant integral
\begin{equation}\label{B.17}
I=\int\left[\mathbf{g}^{ik}\bar{R}_{ik}(W)
+\frac{8\pi}{3}W_i\mathbf{j}^i\right]d\Omega.
\end{equation}
From it, when eq. (\ref{B.6}) is assumed to hold, by means of an
infinitesimal coordinate transformation we get the four identities
\begin{eqnarray}\label{B.18}
-(\mathbf{g}^{is}\bar{R}_{ik}(W)
+\mathbf{g}^{si}\bar{R}_{ki}(W))_{,s}
+\mathbf{g}^{pq}\bar{R}_{pq,k}(W)\\\nonumber
+\frac{8\pi}{3}\mathbf{j}^i(W_{i,k}-W_{k,i})=0.
\end{eqnarray}
This equation can be rewritten as
\begin{eqnarray}\label{B.19}
-2(\mathbf{g}^{(is)}\bar{R}_{(ik)}(\Gamma))_{,s}
+\mathbf{g}^{(pq)}\bar{R}_{(pq),k}(\Gamma)\\\nonumber
=2\mathbf{g}^{[is]}_{~,s}\bar{R}_{[ik]}(\Gamma) +\mathbf{g}^{[is]}
\left\{\bar{R}_{[ik],s}(\Gamma) +\bar{R}_{[ks],i}(\Gamma)
+\bar{R}_{[si],k}(\Gamma)\right\}
\end{eqnarray}
where the redundant variable $W^i_{kl}$ no longer appears. The
metric tensor $s_{ik}$ is defined by equation (\ref{1.2}), and
just for the tensor $T_{ik}$ we shall make an exception to the
general rule that prevails in the Einstein-Schr\"odinger theory,
and use $s^{ik}$ and $s_{ik}$ to raise and lower indices,
$\sqrt{-s}$ to produce tensor densities out of tensors. We define
then
\begin{equation}\label{B.20}
\mathbf{T}^{ik}=\sqrt{-s}s^{ip}s^{kq}T_{pq}
\end{equation}
and the weak identities (\ref{B.19}), when all the field equations
hold, will take the form
\begin{equation}\label{B.21}
\mathbf{T}^{ls}_{;s}=\frac{1}{2}s^{lk}
(\mathbf{j}^i\bar{R}_{[ki]}(\Gamma) +K_{iks}\mathbf{g}^{[si]}),
\end{equation}
where the semicolon means covariant derivative with respect to the
Christoffel affine connection
\begin{equation}\label{B.22}
\left\{^{~i}_{k~l}\right\}
=\frac{1}{2}s^{im}(s_{mk,l}+s_{ml,k}-s_{kl,m})
\end{equation}
built with $s_{ik}$. As far as one knows, only the choice
(\ref{1.2}) of the metric and the way of appending sources adopted
in eqs. (\ref{B.11}), (\ref{B.8}), (\ref{B.15}) and (\ref{B.16})
allows rewriting the identities (\ref{B.19}) in so simple and so
physically suggestive a way. The previous impression is
strengthened by eq. (\ref{B.21}): the theory, built in terms of a
non-Riemannian geometry, appears to entail a gravoelectrodynamics
in a dynamically polarized Riemannian spacetime, for which
$s_{ik}$ is the metric, where the two conserved currents ${\mathbf
j}^i$ and $K_{iks}$ are coupled \`a la Lorentz to $\bar{R}_{[ki]}$
and to $\mathbf{g}^{[si]}$ respectively. Two versions of this
gravoelectrodynamics are possible, according to whether
$\mathbf{g}^{ik}$ is chosen to be either a real nonsymmetric or a
complex Hermitian tensor density, like we presently do. The
constitutive relation between electromagnetic inductions and
fields is governed by the field equations in a quite novel and
subtle way: the link between ${\mathbf g}^{[ik]}$ and
$\bar{R}_{[ik]}$ is not the simple algebraic one usually
attributed to the vacuum, with some metric that raises or lowers
indices, and builds densities from tensors. It is a differential
one, and a glance to the field equations suffices to become
convinced that understanding its properties is impossible without
first finding and perusing the exact solutions of the theory.\par

\section{Solutions of the Hermitian theory that depend on three coordinates}\label{C}
 We assume that Greek indices take the values 1,2 and 4,
while Latin indices run from 1 to 4. Let the real symmetric tensor
$h_{ik}$ be the metric for a vacuum solution to the field
equations of the general relativity of 1915, which depends on the
three co-ordinates $x^{\lambda}$, not necessarily all spatial in
character, and for which $h_{\lambda 3}=0$. We consider also an
antisymmetric purely imaginary tensor $a_{ik}$, which depends too
only on the co-ordinates  $x^{\lambda}$, and we assume that its
only nonvanishing components are $a_{\mu 3}=-a_{3 \mu}$. Then we
form the mixed tensor
\begin{equation}\label{C.1}
\alpha_i^{~k}=a_{il}h^{kl}=-\alpha^k_{~i},
\end{equation}
where $h^{ik}$ is the inverse of $h_{ik}$, and we define the
Hermitian fundamental form $g_{ik}$ as follows:
\begin{eqnarray}\nonumber
g_{\lambda\mu}=h_{\lambda\mu},\\\label{C.2}
g_{3\mu}=\alpha_3^{~\nu}h_{\mu\nu},\\\nonumber
g_{33}=h_{33}-\alpha_3^{~\mu}\alpha_3^{~\nu}h_{\mu\nu}.
\end{eqnarray}
When the three additional conditions
\begin{equation}\label{C.3}
\alpha^3_{~\mu,\lambda}-\alpha^3_{~\lambda,\mu}=0
\end{equation}
are fulfilled, the affine connection $\Gamma^i_{kl}$ which solves
eqs. (\ref{A.2}) has the nonvanishing components
\begin{eqnarray}\label{C.4}
\Gamma^{\lambda}_{(\mu\nu)}=\left\{^{~\lambda}_{\mu~\nu}\right\}_{(h)},
\\\nonumber
\Gamma^{\lambda}_{[3\nu]}=\alpha^{~\lambda}_{3~,\nu}
-\left\{^{~3}_{3~\nu}\right\}_{(h)}\alpha^{~\lambda}_3
+\left\{^{~\lambda}_{\rho~\nu}\right\}_{(h)}\alpha^{~\rho}_3,
\\\nonumber
\Gamma^3_{(3\nu)}=\left\{^{~3}_{3~\nu}\right\}_{(h)},
\\\nonumber
\Gamma^{\lambda}_{33}=\left\{^{~\lambda}_{3~3}\right\}_{(h)}
-\alpha^{~\nu}_3\left(\Gamma^{\lambda}_{[3\nu]}
-\alpha^{~\lambda}_3\Gamma^3_{(3\nu)}\right);
\end{eqnarray}
we indicate with $\left\{^{~i}_{k~l}\right\}_{(h)}$ the
Christoffel connection built with $h_{ik}$. We form now the Ricci
tensor (\ref{A.6}). When eqs. (\ref{A.3}), i.e., in our case, the
single equation
\begin{equation}\label{C.5}
(\sqrt{-h}~\alpha^{~\lambda}_3 h^{33})_{,\lambda}=0,
\end{equation}
and the additional conditions, expressed by eqs. (\ref{C.3}), are
satisfied, the components of $R_{ik}(\Gamma)$ can be written as
\begin{eqnarray}\nonumber
R_{\lambda\mu}=H_{\lambda\mu},
\\\label{C.6}
R_{3\mu}=\alpha^{~\nu}_3H_{\mu\nu}+\left(\alpha^{~\nu}_3
\left\{^{~3}_{3~\nu}\right\}_{(h)}\right)_{,\mu},
\\\nonumber
R_{33}=H_{33}-\alpha^{~\mu}_3\alpha^{~\nu}_3H_{\mu\nu},
\end{eqnarray}
where $H_{ik}$ is the Ricci tensor built with
$\left\{^{~i}_{k~l}\right\}_{(h)}$. $H_{ik}$ is zero when the seed
metric $h_{ik}$ is a vacuum solution of the field equations of
general relativity, as supposed; therefore, when eqs. (\ref{C.3})
and (\ref{C.5}) hold, the Ricci tensor, defined by eqs.
(\ref{C.6}), satisfies eqs. (\ref{A.4}) and (\ref{A.5}) of the
Hermitian theory of relativity.\par The task of solving equations
(\ref{A.2})-(\ref{A.5}) reduces, under the circumstances
considered here, to the simpler task of solving eqs. (\ref{C.3})
and (\ref{C.5}) for a given $h_{ik}$.\footnote[7]{This method of
solution obviously applies to Schr\"odin\-ger's purely affine
theory \cite{Schroedinger1948} too.}\par Let us suppose that the
metric tensor is defined by the equation (\ref{1.2}), namely
\begin{equation}\label{C.7}
\sqrt{-s}s^{ik}=\mathbf{g}^{(ik)},
\end{equation}
where $s^{il}s_{kl}=\delta^i_k$ and $s=\det{(s_{ik})}$. When the
fundamental tensor $g_{ik}$ has the form (\ref{C.2}) it is
\begin{equation}\label{C.8}
\sqrt{-g}=\sqrt{-h},
\end{equation}
where $h\equiv\det(h_{ik})$, and
\begin{equation}\label{C.9}
\det{\left(g^{(ik)}\right)}=\frac{1-g^{3\tau}g_{3\tau}}{h}.
\end{equation}
Therefore
\begin{equation}\label{C.10}
\sqrt{-s}=\sqrt{-h}\left(1-g^{3\tau}g_{3\tau}\right)^{1/2},
\end{equation}
hence
\begin{equation}\label{C.11}
s^{ik}=g^{(ik)}\left(1-g^{3\tau}g_{3\tau}\right)^{-1/2}.
\end{equation}
The nonvanishing components of $s_{ik}$ then read
\begin{eqnarray}\nonumber
s_{\lambda\mu}=\left(1-g^{3\tau}g_{3\tau}\right)^{1/2}h_{\lambda\mu}
+\left(1-g^{3\tau}g_{3\tau}\right)^{-1/2}
h_{33}\alpha^3_{~\lambda}\alpha^3_{~\mu},\\\label{C.12}
s_{33}=\left(1-g^{3\tau}g_{3\tau}\right)^{1/2}h_{33},
\end{eqnarray}
and the square of the interval $\dd s^2=s_{ik}\dd x^i\dd x^k$
eventually comes to read
\begin{equation}\label{C.13}
\dd s^2=\left(1-g^{3\tau}g_{3\tau}\right)^{1/2}h_{ik}\dd x^i\dd
x^k
-\left(1-g^{3\tau}g_{3\tau}\right)^{-1/2}h_{33}\left(\dd\xi\right)^2.
\end{equation}
In keeping with (\ref{C.3}), we have defined $\alpha^3_{~\mu}$ as
\begin{equation}\label{C.14}
\alpha^3_{~\mu}=i\xi_{,\mu},
\end{equation}
in terms of the real function $\xi(x^{\lambda})$.

\newpage
\bibliographystyle{amsplain}

\begin{thebibliography}{}


\bibitem{Einstein1925} Einstein, A. (1925). \textit{ S. B. Preuss. Akad. Wiss.},
{\bf 22}, 414.

\bibitem{ES1946} Einstein, A., and Straus, E.G. (1946).  \textit{ Ann. Math.},
{\bf 47}, 731.

\bibitem{Einstein1948} Einstein, A. (1948). \textit{Rev. Mod. Phys.},
{\bf 20}, 35.

\bibitem{EK1955} Einstein, A., and Kaufman, B. (1955).  \textit{Ann. Math.},
{\bf 62}, 128.

\bibitem{Schroedinger1947a} Schr\"odinger, E. (1947).
\textit{Proc. R. I. Acad.}, {\bf 51A}, 163.

\bibitem{Schroedinger1947b} Schr\"odinger, E. (1947).
\textit{Proc. R. I. Acad.}, {\bf 51A}, 205.

\bibitem{Schroedinger1948} Schr\"odinger, E. (1948).
\textit{Proc. R. I. Acad.}, {\bf 52A}, 1.

\bibitem{Schroedinger1951} Schr\"odinger, E. (1951).
\textit{Proc. R. I. Acad.}, {\bf 54A}, 79.

\bibitem{Papapetrou1948} Papapetrou, A. (1948).
\textit{Proc. R. I. Acad.}, {\bf 52A}, 69.

\bibitem{Wyman1950} Wyman, M. (1950). \textit{Can. J. Math.}, {\bf 2}, 427.

\bibitem{Kursunoglu1952a} Kur\c{s}uno\u{g}lu, B. (1952).
\textit{Proc. Phys. Soc. A}, {\bf 65}, 81.

\bibitem{Kursunoglu1952b} Kur\c{s}uno\u{g}lu, B. (1952). \textit{Phys. Rev.},
{\bf 88}, 1369.

\bibitem{Lichnerowicz1954} Lichnerowicz, A., (1954).
\textit{J. Rat. Mech. Anal.}, {\bf 3}, 487.

\bibitem{Hely1954a} H\'ely, J. (1954). \textit{Comptes Rend. Acad. Sci.
(Paris)}, {\bf 239}, 385.

\bibitem{Hely1954b} H\'ely, J. (1954). \textit{Comptes Rend. Acad. Sci.
(Paris)}, {\bf 239}, 747.

\bibitem{Lichnerowicz1955} Lichnerowicz, A., (1955).
\textit{Th\'eories relativistes de la gravitation et de
l'\'e\-lectro\-magn\'etisme}, Masson, Paris.

\bibitem{Tonnelat1955} Tonnelat,
M. A. (1955). \textit{La Th\'eorie du Champ Unifi\'e d'Einstein},
Gauthier-Villars, Paris.

\bibitem{Narlikar Rao1956} Narlikar, V.V., and Rao, B.R. (1956).
\textit{Proc. Nat. Inst. Sci. India} {\bf 21A}, 409.

\bibitem{Treder1957} Treder, H. (1957). \textit{Ann. Phys. (Leipzig)},
{\bf 19}, 369.

\bibitem{Hlavaty1957} Hlavat\'y, V. (1957). \textit{Geometry of Einstein's Unified
Field Theory} Noordhoff, Groningen.

\bibitem{Schroedinger1954} Schr\"odinger, E. (1954). \textit{Space-Time
Structure}, Cambridge University Press, Cambridge.

\bibitem{Borchsenius1978} Borchsenius, K. (1978). \textit{Nuovo
Cimento}, {\bf 46A}, 403.

\bibitem{Antoci1991} Antoci, S. (1991). \textit{Gen. Rel. Grav.},
{\bf 23}, 47. Also: http://arxiv.org/abs/gr-qc/0108052.

\bibitem{Antoci1987} Antoci, S. (1987). \textit{Ann. Phys. (Leipzig)},
{\bf 44}, 297;\par\noindent http://arXiv.org/abs/gr-qc/0108042.

\bibitem{ALM2005} Antoci, S., Liebscher, D.-E. and Mihich, L. (2005).
\textit{Gen. Rel. Grav.}, {\bf 37}, 1191;\par\noindent
http://arxiv.org/abs/gr-qc/0405064.

\bibitem{Antoci2009} Antoci, S. (2009). \textit{Nuovo Cimento} {\bf 124B},
121;\par\noindent http://arxiv.org/abs/0803.3587.

\bibitem{ALM2006} Antoci, S., Liebscher, D.-E. and Mihich, L.
(2008). \textit{Annales Fond. de Broglie}, {\bf 33},
221;\par\noindent http://arxiv.org/abs/gr-qc/0604003.

\bibitem{Schouten1954} Schouten, J.A. (1954). \textit{Ricci-calculus; an
introduction to tensor analysis and its geometrical applications},
Springer, Berlin.

\end{thebibliography}

\end{document}